\documentclass[preprint]{imsart}

\pdfoutput=1

\RequirePackage[OT1]{fontenc}
\RequirePackage{amsthm,amsmath, amssymb}
\RequirePackage[round, longnamesfirst]{natbib}
\RequirePackage[colorlinks,citecolor=blue,urlcolor=blue]{hyperref}
\usepackage{bbm}
\usepackage{subfigure}
\usepackage{graphicx}

\startlocaldefs
\numberwithin{equation}{section}
\theoremstyle{plain}

\endlocaldefs

\graphicspath{{./figures/}}
\begin{document}

\begin{frontmatter}
\title{On Bayesian quantile regression and outliers}
\runtitle{On Bayesian quantile regression and outliers}

\begin{aug}
\author{\fnms{Bruno} \snm{Santos}\ead[label=e1]{bramos@ime.usp.br}}
\and
\author{\fnms{Heleno} \snm{Bolfarine}\ead[label=e2]{hbolfar@ime.usp.br}}

\address{Institute of Mathematics and Statistics\\
University of S\~ao Paulo \\
Rua do Mat\~ao 1010, Cidade Universit\'aria \\
S\~ao Paulo, Brazil \\
\printead{e1,e2}}

\runauthor{Santos and Bolfarine}

\affiliation{Universidade de S\~ao Paulo}

\end{aug}

\begin{abstract}
In this work we discuss the progress of Bayesian quantile regression models since 
their first proposal and we discuss the importance of all parameters involved in 
the inference process. Using a representation of the asymmetric Laplace distribution
as a mixture of a normal and an exponential distribution, we discuss the relevance of 
the presence of a scale parameter to control for the variance in the model. 
Besides that we consider the posterior distribution of the latent variable 
present in the mixture representation to showcase outlying observations given the 
Bayesian quantile regression fits, where we compare the posterior distribution for 
each latent variable with the others. We illustrate these results with simulation 
studies and also with data about Gini indexes in Brazilian states from years with
census information.
\end{abstract}


\tableofcontents
\end{frontmatter}

\section{Introduction}

Quantile regression can no longer be considered an outsider in the regression
analysis framework, as it has been widely studied in the literature and can 
be found in most statistical software these days. This technique was 
introduced by \citet{koenker:78} as a minimization problem, where the 
conditional quantiles of the response variable is the answer. It was
even first coined as ``regression quantiles'', instead of quantile regression,
by the authors. In fact, the second term is the name of the book by 
\citet{koenker:05} which brings several examples of application, while also 
presenting key asymptotic results that, for instance, allow the construction of 
confidence intervals and hypothesis tests using a few different inferential 
procedures. 

First, this frequentist procedure was not attached to any probability 
distribution, as parameter estimation was made possible through 
linear programming algorithms, while inferential methods, such as 
hypothesis tests and confidence intervals could rely on asymptotic results
or bootstrap, for instance. \citet{koenker:99} connected the asymmetric 
Laplace distribution to these models, where they defined a likelihood 
ratio test using the assumption of this distribution.

\citet{yu:01} introduced Bayesian quantile regression models, assuming in 
the likelihood the asymmetric Laplace distribution, but fixing its scale 
parameter equal to one. In this first proposal, they used an improper prior 
distribution for the regression parameters, but the authors showed that 
they still obtained a proper posterior. Later, \citet{kozumi:11} adopted 
a location-scale mixture of the asymmetric Laplace distribution to build 
a more flexible Markov Chain Monte Carlo (MCMC) scheme to draw samples 
from the posterior distribution. \citet{khare:12} proved that this new
sampling algorithm converges at a geometric rate. 

Recently, \citet{sriram:13} demonstrated posterior consistency for quantile
estimates using the assumption of the asymmetric Laplace distribution, as a 
misspecified model. In fact, when building these models for the same dataset 
one considers that for each quantile of interest a different likelihood should 
be properly combined with the prior, to produce a posterior distribution. This 
makes the misspecified model assumption very reasonable. Using a similar idea, 
\citet{yang:15} argue that fixing the $\sigma$ parameter, one needs to make a 
small modification in the posterior covariance matrix of the regression 
parameters, in order to get reliable confidence intervals. Although, we agree 
with the misspecified model result, we discuss here in this paper that 
one should not fix $\sigma$, but instead should learn from its posterior 
distribution.

In the nonparametrics and semiparametric literature, there are also proposals
for Bayesian quantile regression models. For instance, using Dirichlet processes, 
\citet{kottas:01} suggest a model for median regression, while \citet{kottas:09} 
and \citet{taddy:10} study models for all quantiles. Non-crossing quantiles 
planes, which is a concern when dealing with quantile regression, are proposed 
by \citet{reich:10} and \citet{tokdar:11}, considering Bernstein polynomial 
bases and functions of Gaussian processes, respectively. In an interesting way, 
these proposals are able to produce quantile estimates, without relying on the
asymmetric Laplace distribution.

Concerning outlying observations, in the frequentist literature, \citet{santos:15b}
proposed influence measures to identify observations that might affect the 
model fit. They considered the likelihood displacement function to determine 
whether one observation would be deemed influential or not. In the process, 
the model is fit again for every observation, in order to obtain the 
parameter estimates without each point. This could become computationally 
challenging for data with high dimensions. Instead, we propose in this
paper, in the light of the Bayesian model, to compare the posterior 
distribution of the latent variable $v_i$ for each observation, in order
to find those most distant points from the others.

The paper is organized as follows. In Section 2, we give a brief review of 
Bayesian quantile regression, discussing some parameters, which usually do
not receive enough attention in the literature. In Section 3, we propose 
the use of the posterior distribution of the latent variable $v_i$ as a 
measure of distance between the observations, suggesting a possible manner 
to identify outliers in the sample. Moreover, in Section 4, we present two 
simulation studies to check how these proposed methods vary in 
different scenarios, with zero, one or two outliers. We illustrate our proposal 
with an application in Section 5, where we argue about the presence of more than
one outlier in data about the Gini indexes in Brazilian states. We finish with 
our final remarks in Section 6.
  
\section{Bayesian quantile regression}
\label{bayesQR}

In quantile regression models, the interest lies, for example, considering just
linear terms, in the following model
\begin{equation*}
 Q_y(\tau | x) = x^\prime \beta(\tau),
\end{equation*}
which states basically that the $\tau$th conditional quantile of $Y$ given $X$ 
is assumed to follow a linear model with coefficients $\beta(\tau)$. A first 
model to produce such estimates goes back to \citet{koenker:78}, where the 
authors proposed, given a sample of $n$ pairs $(y_i, X_i)$, to minimize the 
following weighted absolute sum 
\begin{equation} \label{minQR}
 \sum_{i=1}^n \rho_\tau(y_i - x_i^\prime \beta),
\end{equation}
where $\rho_\tau(u) = u(\tau - \mathbb{I}(u < 0))$ and $I(.)$ is the indicator
function, through linear programming algorithms.

In the Bayesian paradigm, \citet{yu:01} used the asymmetric Laplace distribution
in the likelihood, with density
\begin{equation*} \label{densALD}
 f(y | \mu, \sigma, \tau) = \frac{\tau(1-\tau)}{\sigma} \exp \left\{ - 
 \rho_\tau \left( \frac{y - \mu}{\sigma} \right) \right\},
\end{equation*}
due to the fact that its location parameter, $\mu \in \mathbb{R}$, is the $\tau$th
quantile of the distribution. In fact, the maximum likelihood estimator when 
we replace $\mu$ for $x^\prime \beta$, matches the estimator obtained by the 
minimization in \eqref{minQR}, for the frequentist model. 

\begin{figure}
\begin{center}
\subfigure[]{
\resizebox*{5.5cm}{!}{\includegraphics{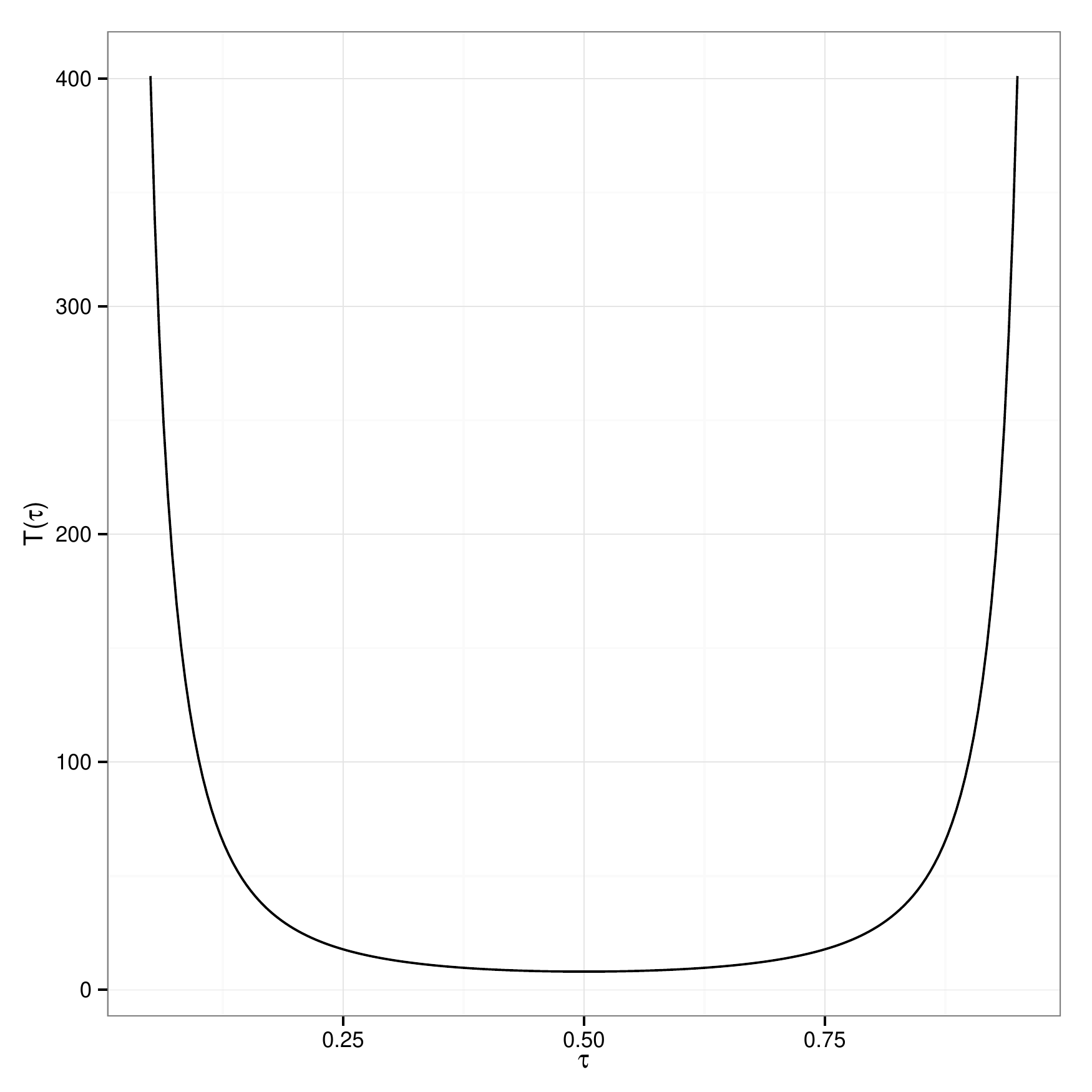}}
\label{figT}}\hspace{5pt}
\subfigure[]{
\resizebox*{5.5cm}{!}{\includegraphics{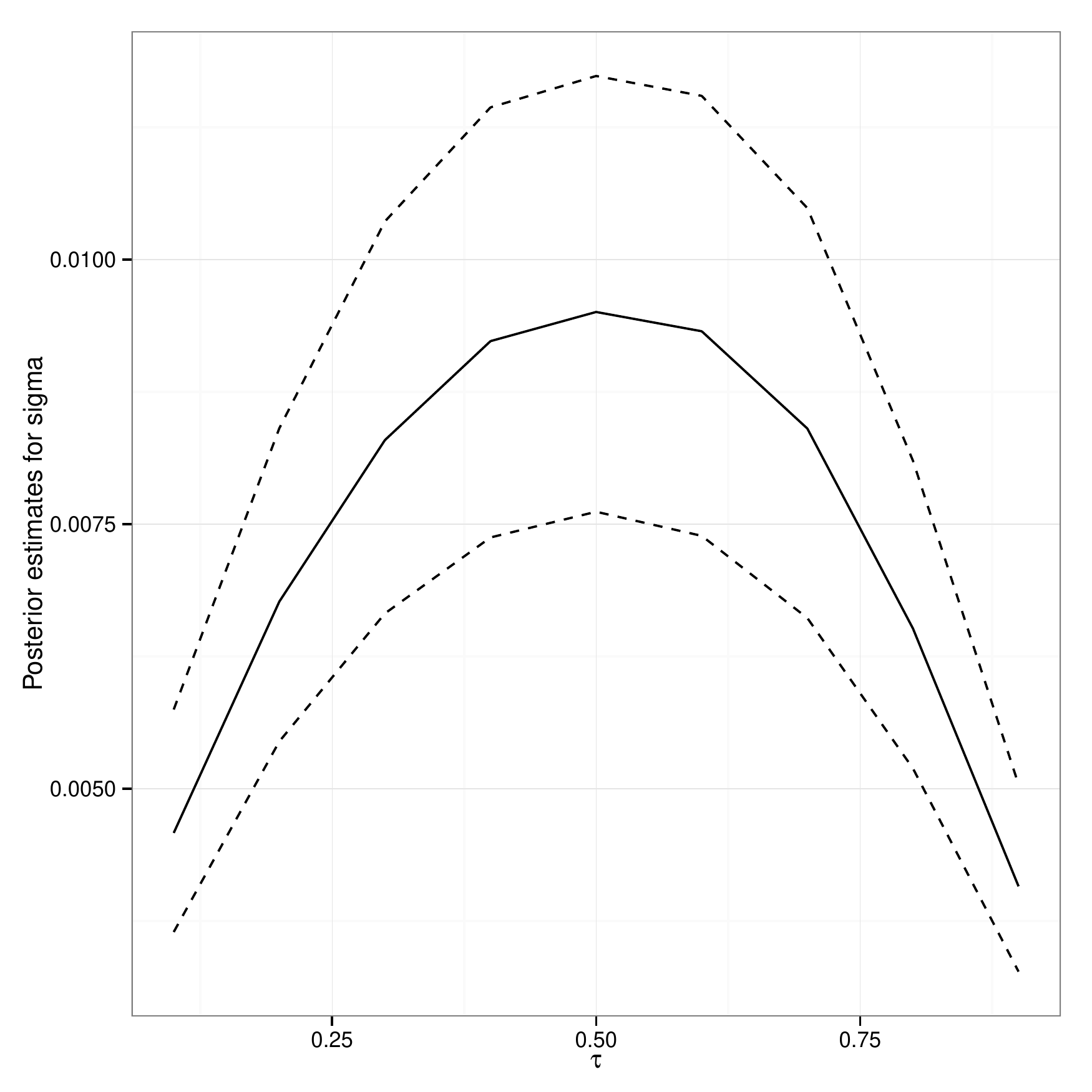}}
\label{figSigma}}
\caption{(a) $T(\tau)$ function which is part of the variance in an asymmetric 
Laplace distribution. (b) Posterior estimates for $\sigma$ in the model analyzed 
in the application section.}
\label{figureSigma}
\end{center}
\end{figure}

Still about the asymmetric Laplace distribution, its mean and variance can be 
written as
\begin{equation*}
 E(Y) = \mu + \frac{\sigma (1-2\tau)}{\tau(1-\tau)}, \quad \mbox{Var}(Y) = \sigma^2 T(\tau),
\end{equation*}
where $\sigma > 0$ is the scale parameter and 
$T(\tau) = (1-2\tau+2\tau^2)/((1-\tau)^2\tau^2)$. The function $T(\tau)$, from 
which depends the variance of $Y$ is presented in Figure~\ref{figT}. One can 
see that this function is U-shaped, so for fixed $\sigma$ the variance is 
greater for smaller or larger quantiles. In their first proposal, \citet{yu:01} 
assumed $\sigma = 1$, automatically increasing the variability for lower and 
greater quantiles, and followed their inference drawing posterior samples for
$\beta(\tau)$.

By giving $\sigma$ a prior distribution, for example, the inverse gamma 
distribution, one can carry on the inference in a more complete way, because
the posterior distribution for $\sigma$ takes into account the data variation
and the variation due to the asymmetric Laplace in the likelihood. For 
instance, in Figure~\ref{figSigma}, we have the mean posterior estimates 
for different quantiles, $\tau = 0.1, 0.2, \ldots, 0.9$, in the application
studied in Section~\ref{secApplication}. It is easy to see that the estimates 
for $\sigma$ adapt according to the quantile and the function $T(\tau)$, and 
that by fixing $\sigma$, which was done by \citet{yu:01} and suggested by 
\citet{yang:15}, one loses such result.

Yet about Bayesian quantile regression, in the modeling framework, 
\citet{kozumi:11} proposed a location-scale mixture representation of the 
asymmetric Laplace distribution, combining a normal distribution conditional 
on an exponential distribution with mean $\sigma$, as follows
\begin{equation*} \label{locationScale}
 Y | v \sim N(\mu + \theta v, \psi^2 \sigma v),
\end{equation*}
where $\theta = (1-2\tau)/(\tau(1-\tau)), \psi^2 = 2/(\tau(1-\tau))$. The 
marginal distribution of $Y$ is the asymmetric Laplace with parameters $\mu$,
$\sigma$ and $\tau$. Now, if we substitute $\mu = x^\prime \beta(\tau)$ and
give a normal prior distribution for $\beta(\tau)$ we have that the full
conditional posterior distribution for the quantile regression parameters is
also normal, making it easier to draw samples from the posterior. In a similar 
way, the full conditional posterior distribution for $\sigma$ is inverse gamma,
if we assume an inverse gamma distribution in the prior. 

Moreover, the latent variable $v_i$, which by construction have an exponential 
prior distribution also needs to be updated in the MCMC algorithm. The full 
conditional posterior distribution for each $v_i$ is proportional to 
\begin{equation} \label{densGIG}
 v_i^{\nu - 1}\exp \left\{ -\frac{1}{2} (\delta_i^2v_i^{-1} + \zeta^2v_i)\right\},
\end{equation}
that is the kernel of a generalized inverse Gaussian distribution. Because each
$v_i$ has its own posterior distribution, that depends on the residual value for 
each observation, this information can be used to compare all observations, even
to identify possible outliers.

All the details of the posterior distributions of all parameters can be found 
in \citet{kozumi:11}.

\section{Outliers observations given the quantile regression fits}
\label{outSection}

Due to the location-scale mixture representation of the asymmetric Laplace, a 
latent variable $v_i$ is added in the modeling scheme for each observation. 
Before updating with data, every $v_i$ is assumed to have an exponential 
distribution with mean $\sigma$, that with the likelihood produces a posterior
distributed according to a generalized inverse Gaussian as in \eqref{densGIG}
with parameters, 
\begin{equation} \label{parametersPost}
 \nu = \frac{1}{2}, \quad 
 \delta_i^2 = \frac{\big(y_i - x_i^\prime \beta(\tau)\big)^2}{\psi^2 \sigma}, 
 \quad 
 \zeta^2 = \frac{2}{\sigma} + \frac{\theta^2}{\psi^2 \sigma}.
\end{equation}

From the parameters in the posterior distribution of $v_i$, just $\delta_i^2$
varies for each observation. And its value is the weighted squared residual of 
the quantile fit. One can see that for larger values of $\delta_i^2$, while the 
other parameters are kept fixed, the posterior distribution of the latent 
variable $v_i$ has a greater expected value. Therefore, more extreme 
observations present a posterior distribution for its latent variable more
distant from zero. 

From empirical evidence, we see that points that have a completely different 
pattern than the one proposed by the model, have their latent variable 
distributed in a region far from the other observations. Given that 
difference, we propose to use that information to label these data points
as possible outliers, i.e., observations that show an extreme pattern that 
can not be explained by the quantile regression model. These points often 
cause bias in the parameter estimates, so it could be discussed even if 
its presence is indeed necessary.

We propose to measure this distance between one observation from the others,
by comparing the posterior distribution of its latent variable in two different
ways. First, we propose to measure the mean probability of the posterior 
conditional latent variable of being greater than the other respective 
latent variables. Second, we use the Kullback-Leibler divergence to assess
the difference between the conditional posterior distributions of latent 
variables based on the MCMC samples.

\subsection{Mean probability posterior}

\begin{figure}[!tb]
\begin{center}
\includegraphics[scale=0.50]{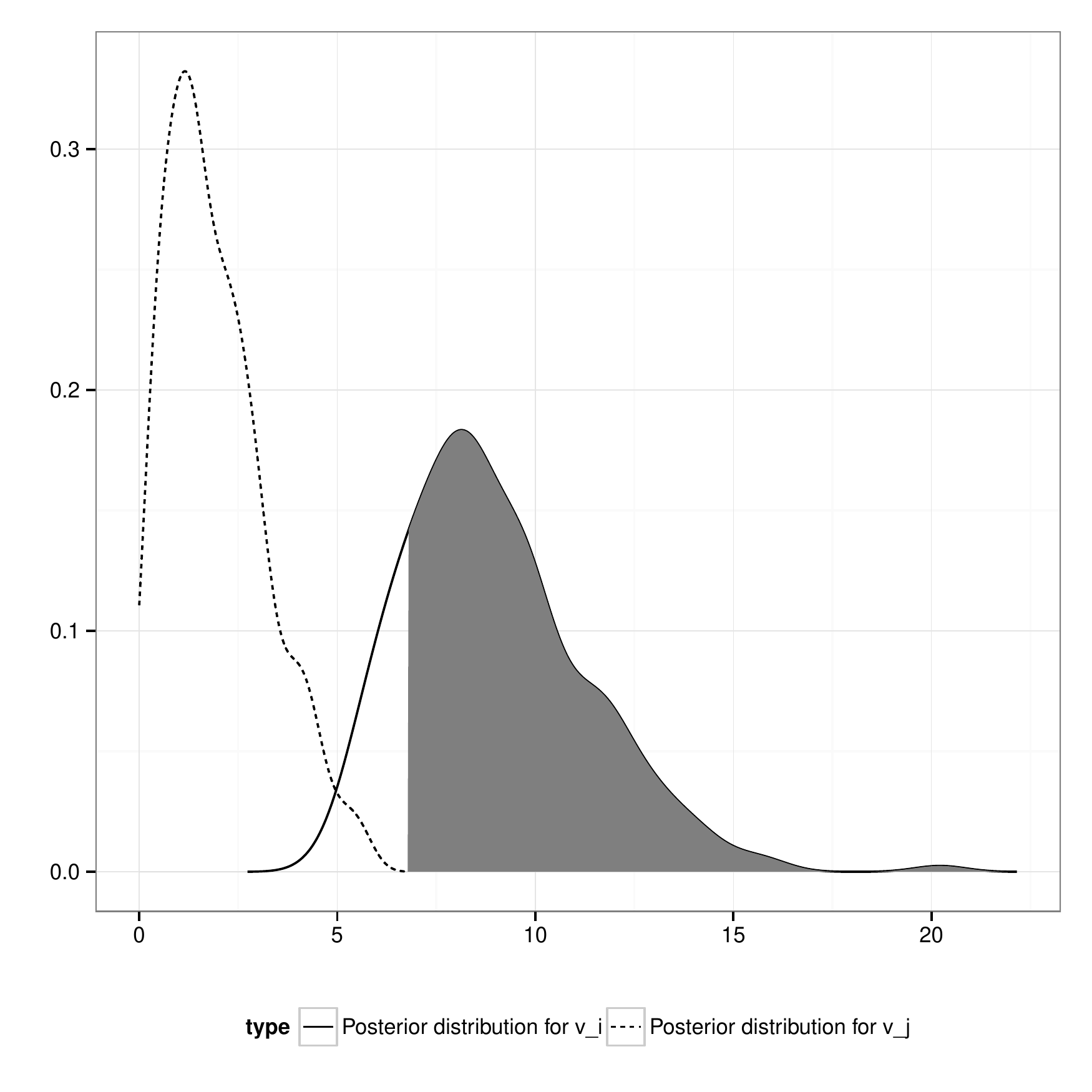}
\caption{Example of how part of the probability in \eqref{probOutlier} 
is calculated, where the area under the dashed line in gray is the 
probability.}
\label{figureProbExample}
\end{center}
\end{figure}

If we define the variable $O_i$, which takes value equal to 1 when the 
$i$th observation is an outlier, and 0 otherwise, then we propose to calculate 
the probability of an observation being an outlier as
\begin{equation} \label{probOutlier}
 P(O_i = 1) = \frac{1}{n-1} \sum_{j \ne i} P(v_i > v_j | \mbox{data}).
\end{equation}
An example of this calculus is presented in Figure~\ref{figureProbExample}, 
where in the probability proposed in \eqref{probOutlier}, we average over 
all observations.

We believe that for points, which are not outliers, this probability should 
be small, possibly close to zero. Given the natural ordering of the residuals,
due to the fact of the posterior parameters depending solely on them as in 
\eqref{parametersPost}, it is expected that some observations present greater
values for this probability in comparison to others. What we think 
that should be deemed as an outlier, ought to be those observations with a
higher $P(O_i = 1)$, and possibly one that is particularly distant from the 
others.

The probability in \eqref{probOutlier} can be approximated given the MCMC 
draws, as follows
\begin{equation*} 
 P(O_i = 1) = \frac{1}{M} \sum_{l=1}^M \mathbb{I}\big(v_i^{(l)} > 
 \max_{k \in 1:M} v_j^{(k)} \big),
\end{equation*}
where $M$ is the size of the chain of $v_i$ after the burn-in period and
$v_i^{(l)}$ is the $l$th draw of this chain.

An important note about this proposal of calculating the probability of an 
observation being an outlier is that this result depends on the quantile, 
therefore a point can be considered an extreme observation for one quantile,
but not the others. This brings more information about the data variation,
as it is more flexible in determining these possible outliers.

\subsection{Kullback-Leibler divergence}

As a second proposal to address these differences between the posterior 
distributions from the distinct latent variables in the model, we 
suggest the use of the Kullback-Leibler divergence proposed by
\citet{kullback:51}, as a more precise method of measuring the 
distance between those latent variables in the Bayesian quantile
regression framework, in its posterior information. This divergence
is defined as 
\begin{equation} \label{kld}
 K(f_i, f_j) = \int \log \left( \frac{f_i(x)}{f_j(x)} \right) f_i(x) dx,
\end{equation}
where in our problem $f_i$ could be the posterior conditional distribution
of $v_i$ and $f_j$ the posterior conditional distribution of $v_j$. Similar 
to the probability proposal in the previous subsection, we should 
average this divergence for one observation based on the distance 
from all others, i.e.,
\begin{equation*}
 KL(f_i) = \frac{1}{n-1} \sum_{j \ne i} K(f_i, f_j)
\end{equation*}

This proposal should be seen as a ratification to the previous probability,
using a more precise measure of distance between the posterior latent 
variables. We expect that when an observation presents a higher value for 
this divergence, it should also present a high probability value of being an
outlier. On one hand, there is the probability value in the range (0, 1), 
which should give some insight of whether one observation should be regarded
as too extreme. On the other hand, there is the Kullback-Leibler, a positive 
valued measure, that could always be analyzed relatively among the observations,
i.e., instead of using its absolute, one could compare how many times this value
is greater than the others. This approach could be helpful to identify 
observations that, for instance, show a not so high probability value, but still
are distributed, in its relative posterior conditional distribution, far from 
the others.

Here, based on the MCMC draws from the posterior of each latent variable, we
estimate the densities in \eqref{kld} using a normal kernel and we compute the 
integral using the trapezoidal rule.

\section{Simulation studies}

In this section, we propose two simulation studies in order to understand 
how these measures defined in the previous section vary according to 
the presence or not of an outlying observation in the case with multivariate
explanatory variables. In the first study, we study the distribution of the 
probability of being an outlier in the absence of such observation. Following,
we discuss the results of the case when there are more than one outlier, 
showing results both for the probability as for the Kullback-Leibler divergence
measure.

\subsection{Simulation 1}

In this first simulation, we try to deal with the scenario where there are no 
outliers, in order to learn the distribution for the probability of being 
an outlier in these situations. We do not present summaries for the 
Kullback-Leibler divergence, as this quantity is not limited and its 
distribution is dependent on other parameters, such as the quantile regression 
parameter, $\sigma$ and the quantile of interest.

We consider the following linear model 
\begin{equation*}
 Y_i = \beta_0 + \beta_1 x_{1i} + \beta_2 x_{2i} + \beta_3 x_{3i} + \epsilon_i, 
 \quad i = 1, \ldots, n,
\end{equation*}
where we set $\beta_0 = 0$, $\beta_1 = 1$, $\beta_2 = -1$, $\beta_3 = 2$,
$\epsilon_i \sim N(0, 4)$, and we draw the three explanatory variables from an
uniform distribution between 0 and 10. We use two samples sizes in this 
study, $n = 100, \, 300$. Each sample was replicated 250 times to produce the
summaries that we discuss next. And three different quantiles were estimated, 
$\tau = 0.25, 0.50, 0.75$.

\begin{figure}
\begin{center}
\subfigure[]{
\resizebox*{5.5cm}{!}{\includegraphics{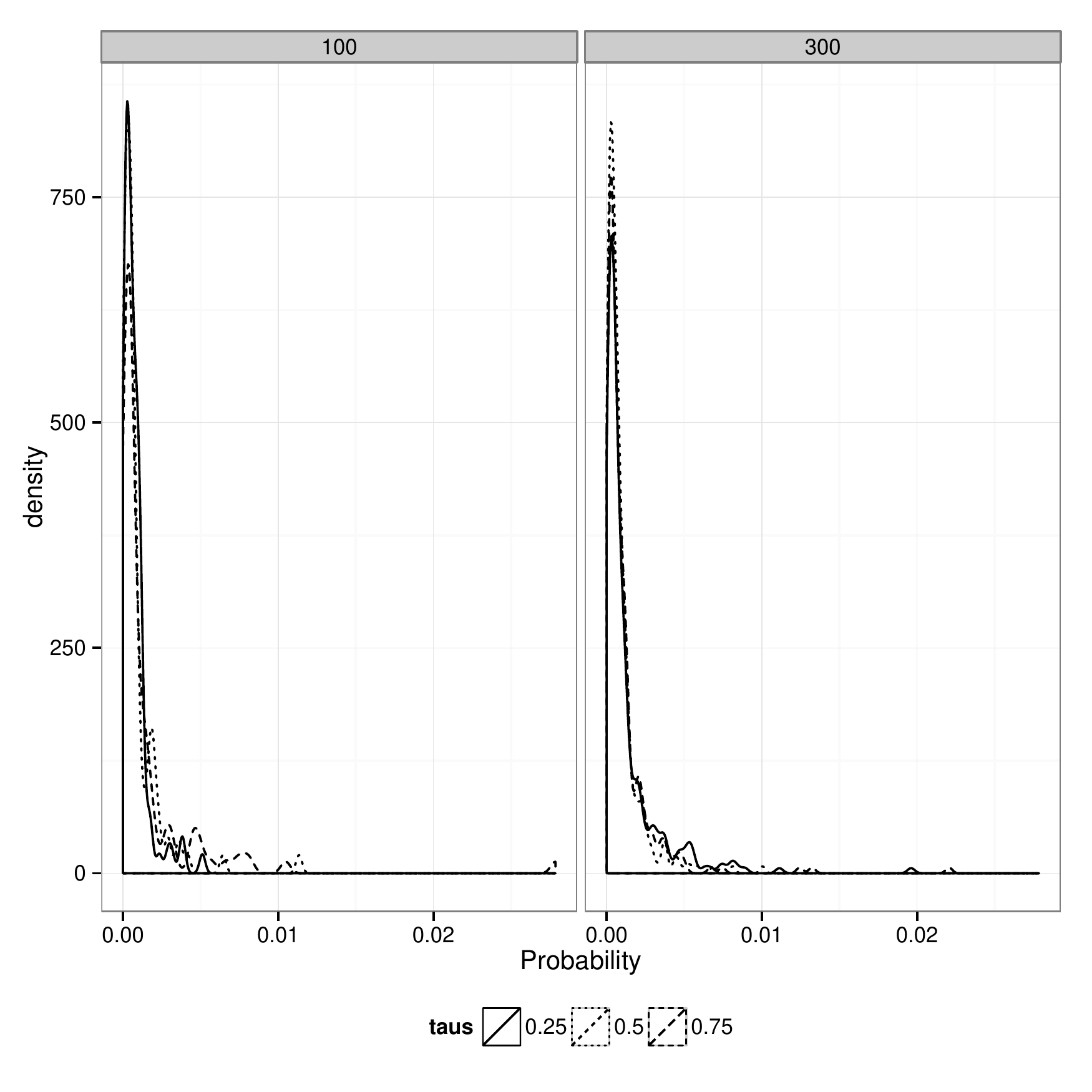}}
\label{prob0a}}\hspace{5pt}
\subfigure[]{
\resizebox*{5.5cm}{!}{\includegraphics{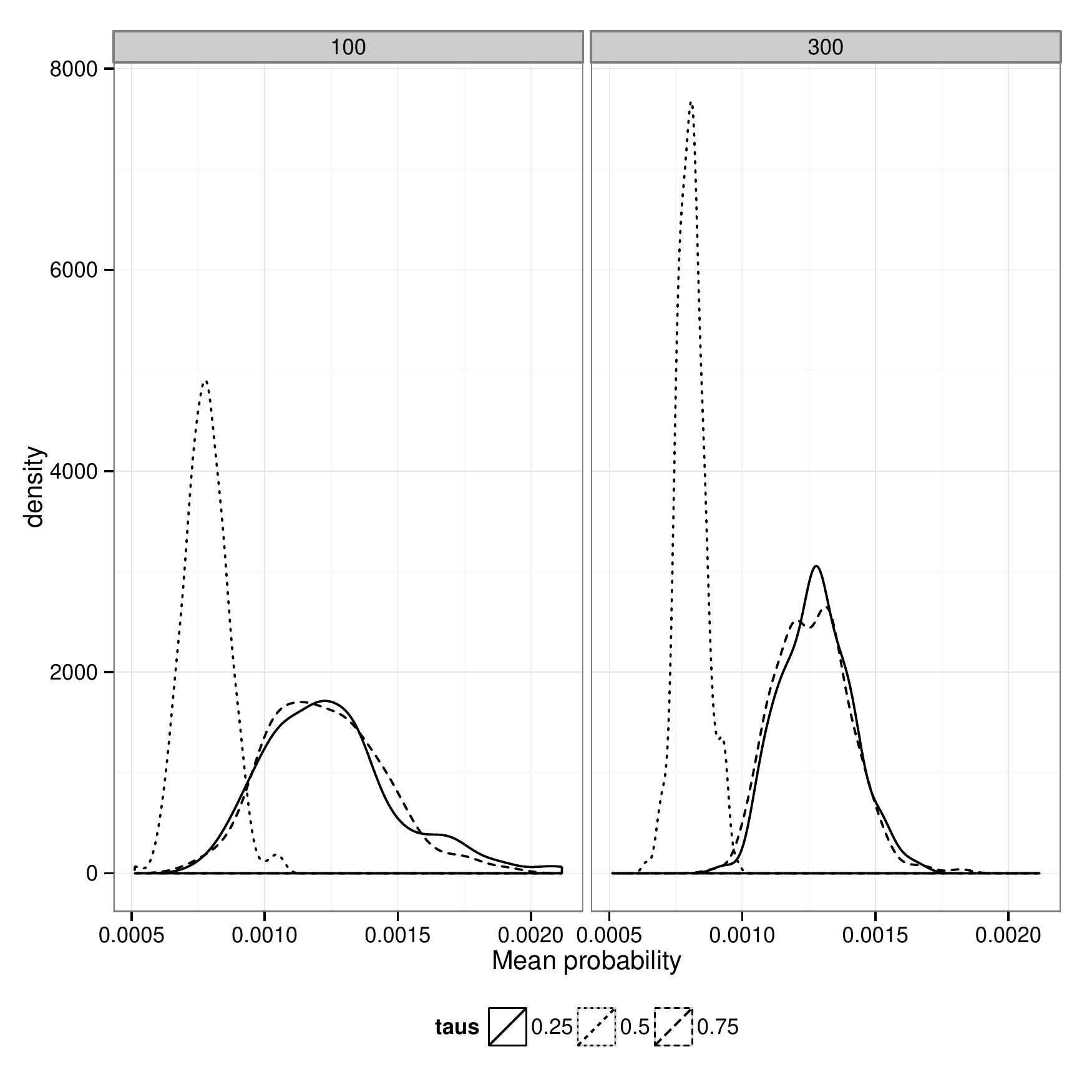}}
\label{prob0b}}
\caption{(a) Distribution of the probabilities for all observations in a randomly 
selected replication in the simulation study. (b) Distribution of the mean 
probability for each replication.}
\label{figureProb0}
\end{center}
\end{figure}

The results for this study can be seen in Figure~\ref{figureProb0}. In the left
part of the figure, we show the probabilities for one of the replications, which
was randomly selected, and where we can see that the probability varies between 
0 and 0.03, approximately. As expected, with the absence of extreme observations,
one should not expect greater values for this probability, as the posterior 
distribution of all latent variables should be relatively close, given that 
residuals should be rather small as well.

If we compare the summaries of all these probabilities in each replication, we get 
the distribution shown in Figure~\ref{prob0b}. Between the different quantiles, the 
probabilities in the conditional median presented smaller mean values in comparison 
with the 0.25th and 0.75th quantiles. For the different sample sizes, the probabilities
decrease slightly as we increase the sample size.  

Given the results in this simulation study, we suggest as a rule of thumb to consider 
outlying observations the ones with non-negligible probability values, possibly greater than 
0.10 at least.

\subsection{Simulation 2}

For this second simulation study, we add until two outliers and record 
both measures to study the presence of outlying observations, while we
replicate these scenarios 250 times as well. We are interested in 
checking the influence of one outlier on the other, when both are present
in the model. We verify this by analyzing the results with just one of these
observations separately and then with both of these in the model. We use 
the same setup as in the previous simulation study, but only considering 
the sample size equal to 100. 

The two outlier observations have the following values for the response 
variables and their respective explanatory variables
\begin{align*}
 y^\ast  &= 30, \quad x_1^\ast = \bar{x}_1, \quad x_2^\ast = 20, \quad x_3^\ast = \bar{x}_3  \\
 y^\star &= 0, \quad x_1^\star = 20, \quad x_2^\star = \bar{x}_2, \quad x_3^\star = \bar{x}_3,
\end{align*}
where $\bar{x}_i$ represents the mean for the $i$th explanatory variable without any 
possible outlier. We argue that $y^\ast$ should be considered an outlier because both 
the response variable value and $x_2^\ast$ are definitely a lot greater than expected, 
specially given the fact that the coefficient for $x_2$ is negative and all other 
observations for the predictor are drawn from an uniform distribution from 0 to 10.
Moreover, for similar reasons $y^\star$ also should be defined as an outlier, as 
$x_1^\star$ is outside the range (0,10) and it produces a response variable smaller 
than expected.

In the following summary results, we use the setup presented in 
Table~\ref{tableSimulation}, where $\times$ represents the presence
of the extreme observation in the scenario.
\begin{table}[!htb]
\centering
\caption{Setup for the different scenarios in Simulation study 2.}
\label{tableSimulation}
\begin{tabular}{ccc}
\cline{2-3}
 & \multicolumn{2}{c}{Outlier} \\
  \cline{2-3}
 & $\ast$ & $\star$ \\
\hline
Scenario 1 & & \\
Scenario 2 & & $\times$ \\
Scenario 3 & $\times$ & $\times$ \\
Scenario 4 & $\times$ & \\
\hline
\hline
\end{tabular}
\end{table}

The summaries for the probabilities in all scenarios are presented in 
Table~\ref{tableSummaries}. It is easy to see when each outlier is added 
separately in the model then their respective probability is always 
high, greater than 0.40 on average. For most scenarios, $y^\ast$ always 
presents a greater probability value in comparison with $y^\star$. 
For both outliers, the probability decreases in the presence of the other,
but still show values far from zero. Overall, these probabilities 
are smaller for quantile 0.5. 

\begin{table}
\centering
\caption{Summary results for the probabilities in each scenario.}
\label{tableSummaries}
\begin{tabular}{ccccccccccc}
\cline{3-6} \cline{8-11}
& & \multicolumn{4}{c}{Outlier $\ast$} & & \multicolumn{4}{c}{Outlier $\star$} \\
\cline{3-6} \cline{8-11}
 $\tau$ & Scenario & Mean & Median & 2.5\% & 97.5\% & & Mean & Median & 2.5\% & 97.5\% \\
\hline
	& 2 & & & & & & 0.505 & 0.488 & 0.228 & 0.783 \\
 $0.1$  & 3 & 0.981 & 0.984 & 0.957 & 0.994 & & 0.452 & 0.436 & 0.197 & 0.721 \\
	& 4 & 1.000 & 1.000 & 0.998 & 1.000 & & & & \\
  & & & & & & & & & \\
	& 2 & & & & & & 0.433 & 0.431 & 0.265 & 0.631 \\
 $0.5$  & 3 & 0.656 & 0.657 & 0.506 & 0.809 & & 0.273 & 0.266 & 0.159 & 0.419 \\
	& 4 & 0.780 & 0.781 & 0.636 & 0.914 & & & & \\
  & & & & & & & & & \\
	& 2 & & & & & & 0.987 & 0.992 & 0.948 & 1.000 \\
 $0.9$  & 3 & 0.810 & 0.823 & 0.596 & 0.935 & & 0.765 & 0.778 & 0.543 & 0.911 \\
	& 4 & 0.841 & 0.852 & 0.656 & 0.961 & & & & \\
\hline
\hline
\end{tabular}
\end{table}

In a interesting way, when we look for the Kullback-Leibler divergences, we 
have an opposite outcome, as we see the greater disparities in the models 
for the conditional median. In Table~\ref{tableSummaries2}, we show the 
mean relative Kullback-Leibler divergence for both outliers, i.e., the 
mean ratio between the divergence between the outliers and a randomly 
selected observation in the sample. We used the comparison with just one 
observation due to the computation burden to calculate for all observations,
but also because we believe that between all observations, which are not 
extreme, the difference would be small. In general, we see that these ratios
are always greater than 9, on average, approximately. In other words, we can 
say that, these outliers show a Kullback-Leibler divergence at least 9 times 
the divergence from a non-outlier observation. 

\begin{table}
\centering
\caption{Summary results for the mean relative Kullback-Leibler divergence in 
each scenario.}
\label{tableSummaries2}
\begin{tabular}{ccccccccccc}
\cline{3-6} \cline{8-11}
& & \multicolumn{4}{c}{Outlier $\ast$} & & \multicolumn{4}{c}{Outlier $\star$} \\
\cline{3-6} \cline{8-11}
 $\tau$ & Scenario & Mean & Median & 2.5\% & 97.5\% & & Mean & Median & 2.5\% & 97.5\% \\
\hline
	& 2 & & & & & & 11.056 & 10.436 & 3.259 & 20.847 \\
 $0.1$  & 3 & 13.988 & 14.880 & 5.968 & 21.908 & & 9.614 & 9.248 & 3.084 & 17.532 \\
	& 4 & 9.125 & 9.380 & 3.321 & 16.125 & & & & \\
  & & & & & & & & & \\
	& 2 & & & & & & 26.343 & 26.871 & 10.636 & 38.251 \\
 $0.5$  & 3 & 28.581 & 29.704 & 15.871 & 38.061 & & 17.153 & 17.365 & 8.042 & 25.147\\
	& 4 & 35.511 & 36.723 & 15.151 & 48.379 & & & & \\
  & & & & & & & & & \\
	& 2 & & & & & & 10.616 & 10.926 & 4.120 & 18.575 \\
 $0.9$  & 3 & 14.882 & 15.448 & 7.055 & 23.127 & & 14.403 & 14.913 & 7.104 & 22.025 \\
	& 4 & 17.217 & 18.383 & 7.424 & 28.111 & & & & \\
\hline
\hline
\end{tabular}
\end{table}

Another interesting aspect of these measures is how they give different conclusions
in respect to these two outliers, $y^\ast$ and $y^\star$. For instance, in the 
0.1th quantile, in the models only with one outlier, the probability is greater for
$y^\star$, while the Kullback-Leibler divergence presents higher values for the
$y^\ast$. On the other hand, in the 0.9th quantile the Kullback-Leibler divergence 
is greater for $y^\star$, even though $y^\ast$ presents higher values of being an
outlier.

\begin{figure}[!tb]
\begin{center}
\includegraphics[scale=0.50]{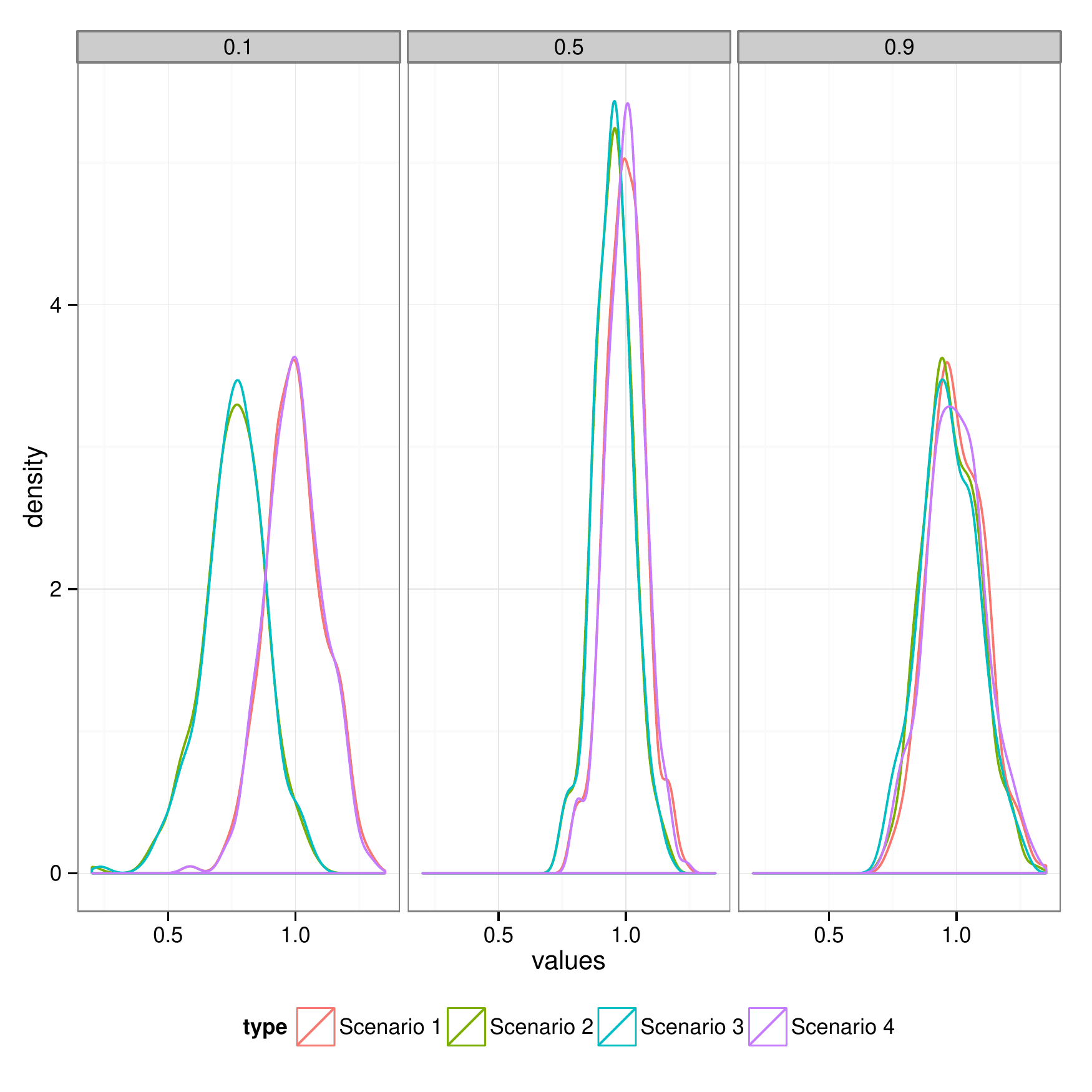}
\caption{Distribution of $\hat{\beta}_1(\tau)$ for $\tau$ = \{0.1, 0.5, 0.9\}.}
\label{figureBeta1}
\end{center}
\end{figure}

\begin{figure}[!tb]
\begin{center}
\includegraphics[scale=0.50]{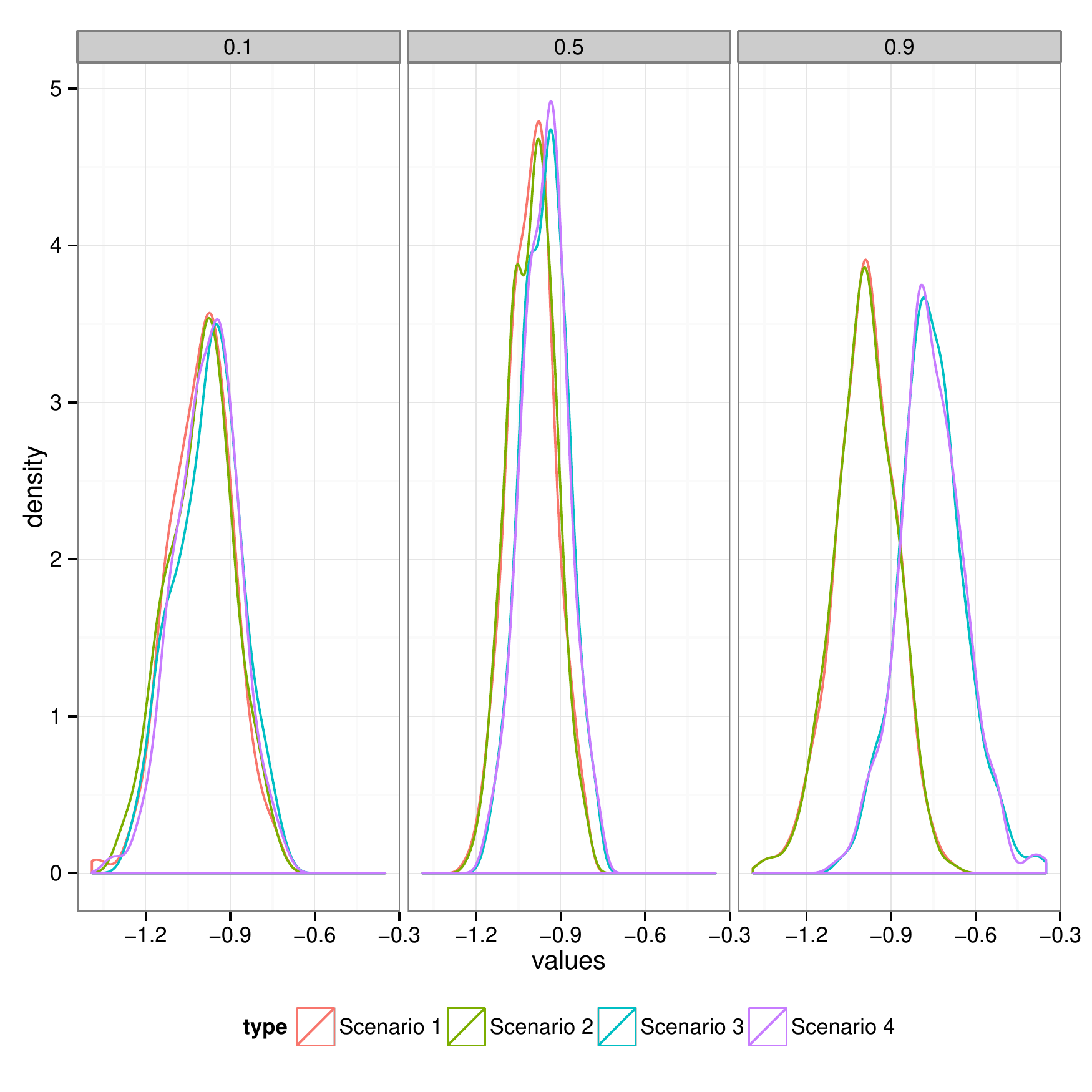}
\caption{Distribution of $\hat{\beta}_2(\tau)$ for $\tau$ = \{0.1, 0.5, 0.9\}.}
\label{figureBeta2}
\end{center}
\end{figure}

Moreover, we present the distribution of the estimates for $\beta_1(\tau)$ and
$\beta_2(\tau)$ in Figure~\ref{figureBeta1} and Figure~\ref{figureBeta2}, 
respectively. For $\beta_1(\tau)$, we can see that its estimates are only 
influenced by the presence of $y^\star$, in Scenarios 2 and 3. And even then
just in the lower quantiles, for instance, the 0.10th quantile.

On the other hand, for $\beta_2(\tau)$, we have that the presence of $y^\ast$
adds a bias in its estimates for greater quantiles, only when this outlier is 
present in Scenario 4, but also when both outliers are present in Scenario 3.

For $\beta_3(\tau)$, we found that neither outlier presented a challenge in its
estimates, as for all scenarios the distribution of $\hat{\beta}_3(\tau)$ was 
not affected by those observations.

\section{Application}
\label{secApplication}

In the interest of using Bayesian quantile regression models to analyze possible
outlying observations, we consider data about Gini indexes in Brazilian states in the 
years 1991, 2000 and 2010, when censuses were conducted countrywide. This data 
comprises the information about 26 states and the Federal District, where the 
Brazilian capital is located, completing 81 observations. 

\begin{figure}
\begin{center}
\subfigure[]{
\resizebox*{5.5cm}{!}{\includegraphics{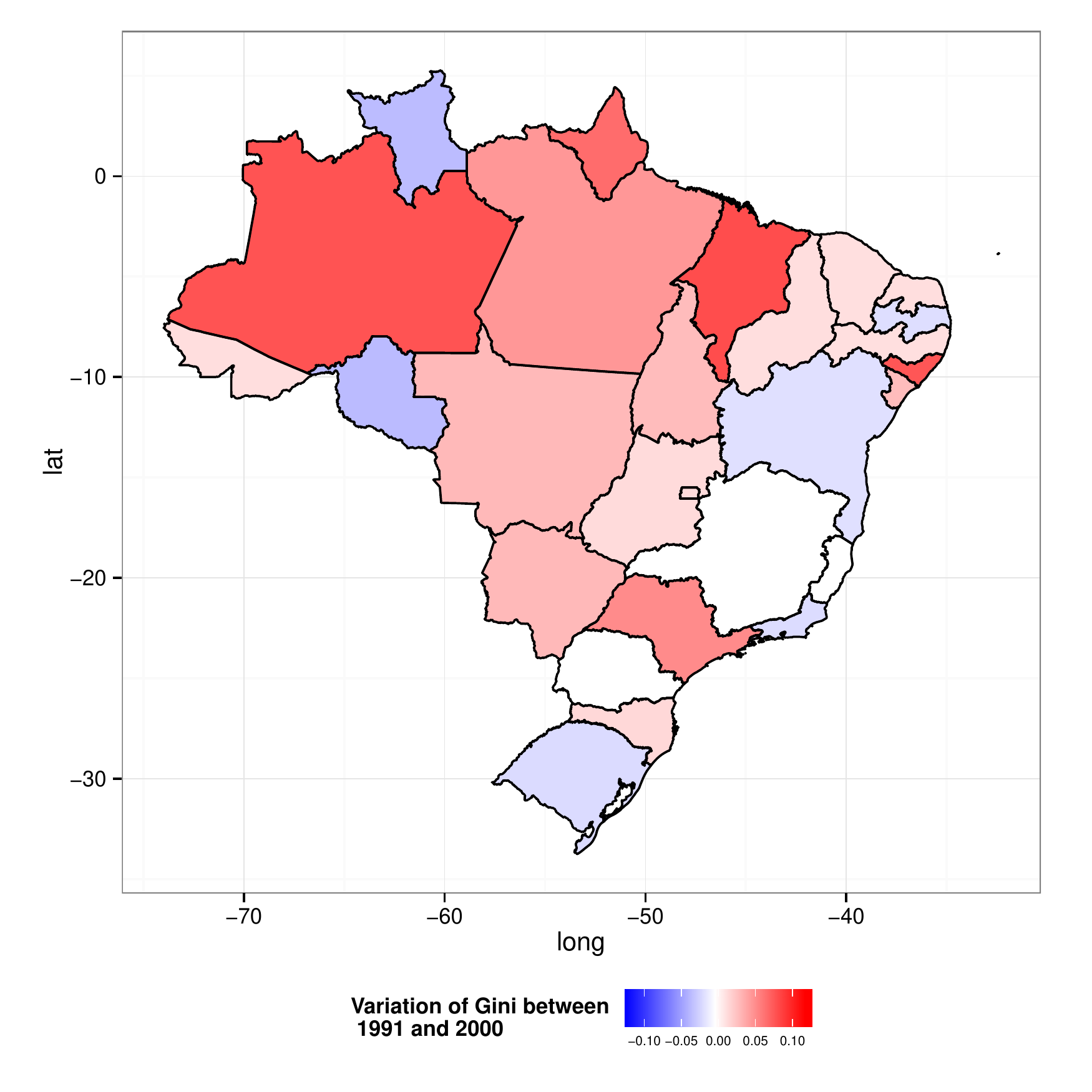}}
\label{fig10a}}\hspace{5pt}
\subfigure[]{
\resizebox*{5.5cm}{!}{\includegraphics{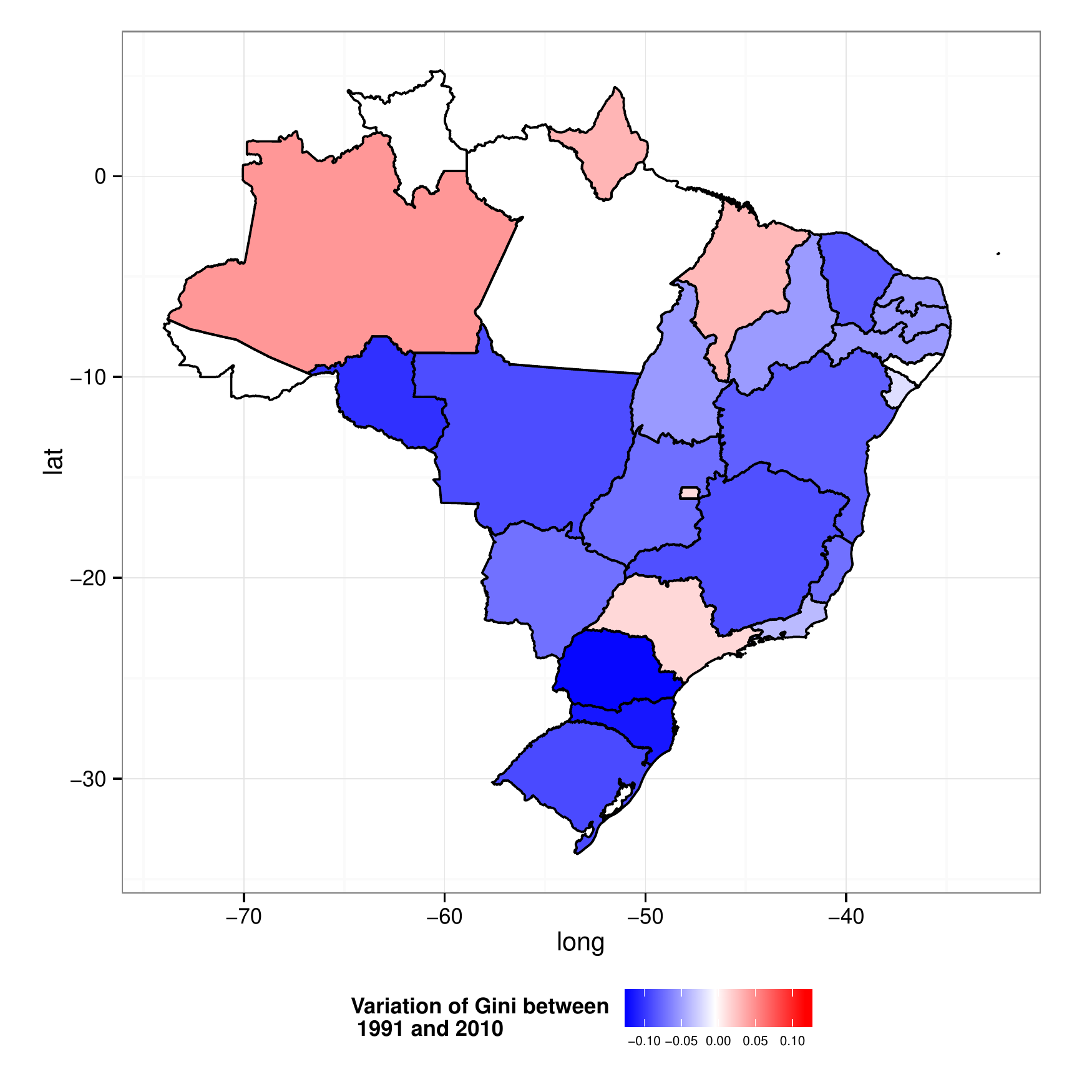}}
\label{fig10b}}
\caption{Variation for the Gini index in the 26 Brazilian states and the Federal 
District, in comparison with 1991 (a) 2000, (b) 2010.}
\label{figureBrazil}
\end{center}
\end{figure}

If we consider data about the whole country, Brazil is usually regarded as a 
highly unequal country, when compared to European countries, for instance. 
Using the Gini index, which gives an indicator of the income inequality, one 
can see that, at least, there was an advance between 1991 and 2010, when this 
measure decreased for several states, as depicted in Figure~\ref{figureBrazil}, 
in spite of the increase for some states at first in 2000.

The following model was proposed to study the conditional quantiles of the Gini index, 
\begin{equation} \label{modelApplication}
 Q_{Y_i}(\tau | x_i) = \beta_0(\tau) + \beta_1(\tau) \mbox{EDUC}_i + 
  \beta_2(\tau) \mbox{INCPC}_i + \beta_3(\tau) \mbox{Y2000}_i +
  \beta_4(\tau) \mbox{Y2010}_i
\end{equation}
where EDUC is the average years of education and INCPC is the income per capita of 
each state, and two indicator variables were used to control for the difference between
the three years, using 1991 as reference. We decided not to transform the response 
variable, the Gini index, which is a number between 0 and 1, as suggested by 
\citet{santos:15}, because even at the most extreme quantiles, the conditional 
estimates were far from the boundaries 0 and 1.

\begin{figure}[!tb]
\begin{center}
\includegraphics[scale=0.50]{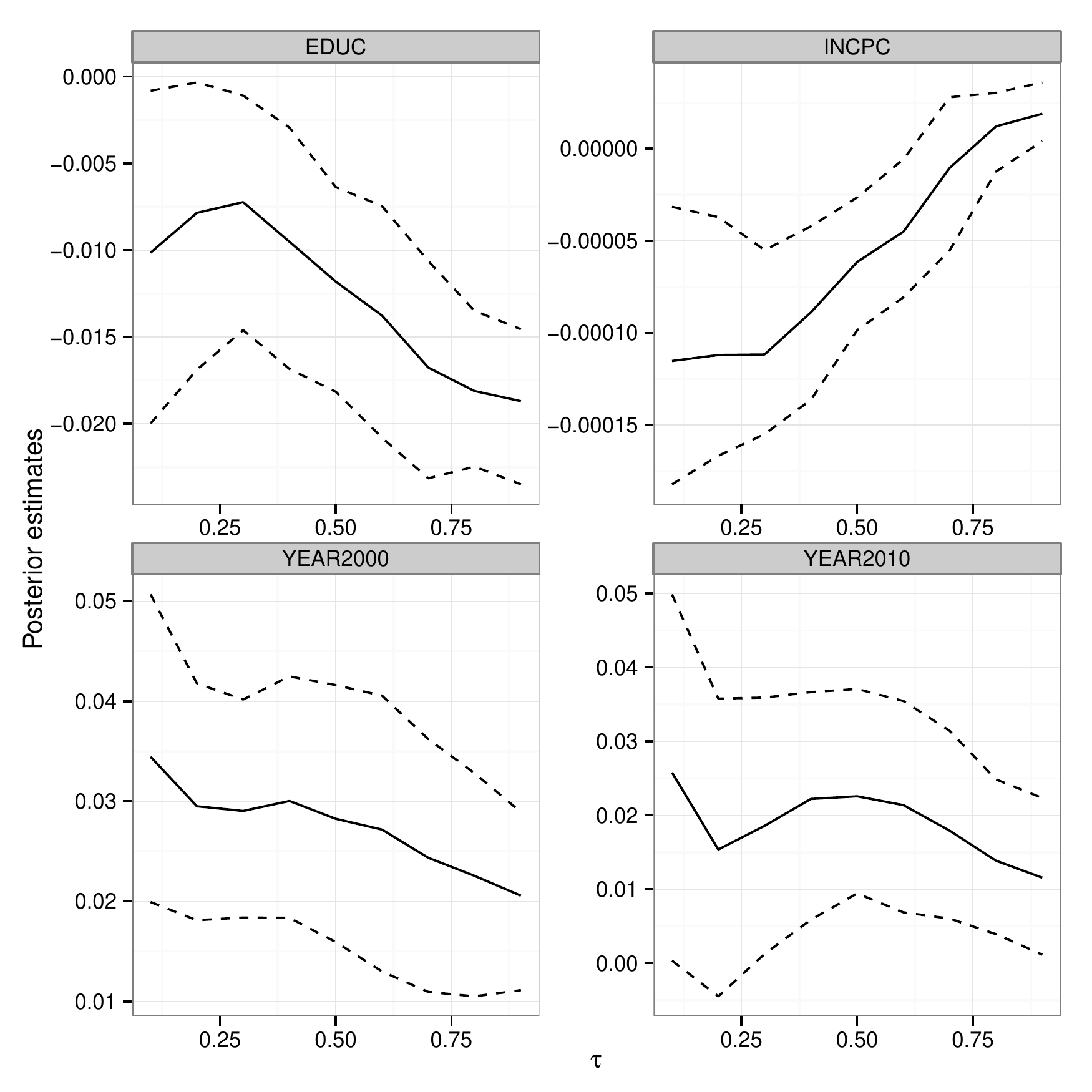}
\caption{Posterior estimates for the quantile regression parameters proposed in the 
model \eqref{modelApplication}.}
\label{figureEstModel}
\end{center}
\end{figure}

The posterior estimates were considered using a chain of size 3000, discarding 
the first 1000 as burn-in. We used a normal distribution $N(0, 100 I)$ for 
$\beta(\tau)$, where $I$ stands for the identity matrix. For $\sigma$, we 
adopted $IG(3/2, 0.1/2)$. The posterior mean and its respective credible interval 
for $\sigma$ in the different quantiles can be seen in Figure~\ref{figSigma}, 
where we can clearly see that the shape of posterior estimates, along with its
credible intervals, have the inverse form of the function $T(\tau)$, presented in
Section~\ref{bayesQR}. Given these results, we defend the importance of using 
a prior distribution for $\sigma$, instead of fixing its value, arguing that the 
posterior distribution naturally adapts to the different sources of variation in 
the modeling process. The posterior mean and 95\% credible intervals for 
$\beta(\tau)$ is presented in Figure~\ref{figureEstModel}.

For years of education, the estimates for $\beta_1(\tau)$ are negative for all 
quantiles, 
but with greater absolute values for $\tau$'s closer to 1. For income per capita, the 
estimates for its respective parameter are also negative, but not significant for 
greater quantiles, $\tau \geqslant 0.6$. Both variables for years presented similar 
estimates with values decreasing along the quantiles, despite having a different 
evolution as shown in Figure~\ref{figureBrazil}. Controlling for other variables, 
we estimate that the Gini indexes in the year 2000 and 2010 in comparison with 1991 
are greater, with this difference being smaller for greater quantiles.

\begin{figure}[!tb]
\begin{center}
\includegraphics[scale=0.50]{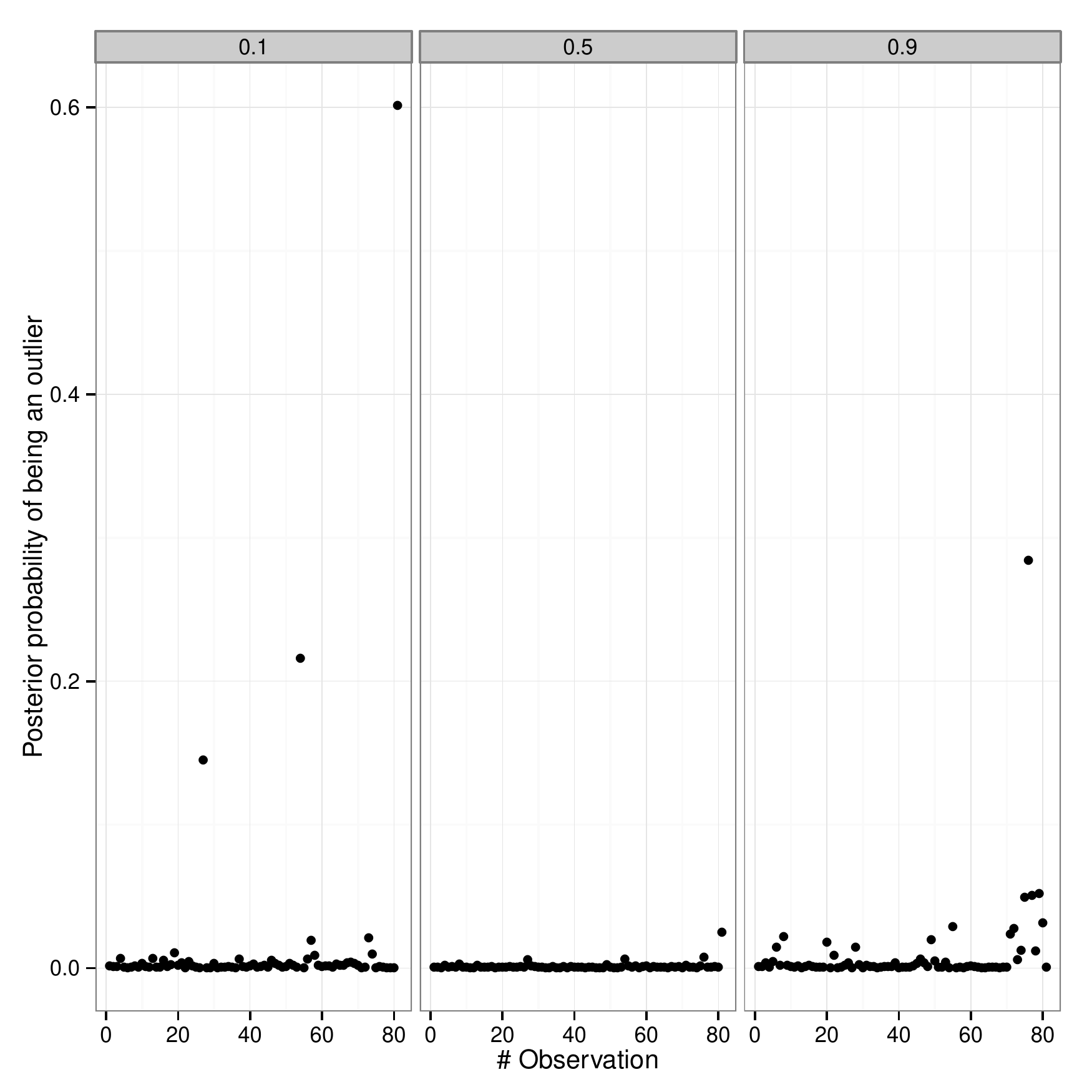}
\caption{Probabilities of being an outlier for $\tau = \{0.1, 0.5, 0.9\}$, 
considering the model in \eqref{modelApplication}}
\label{figureOutliers}
\end{center}
\end{figure}

\begin{figure}[!tb]
\begin{center}
\includegraphics[scale=0.50]{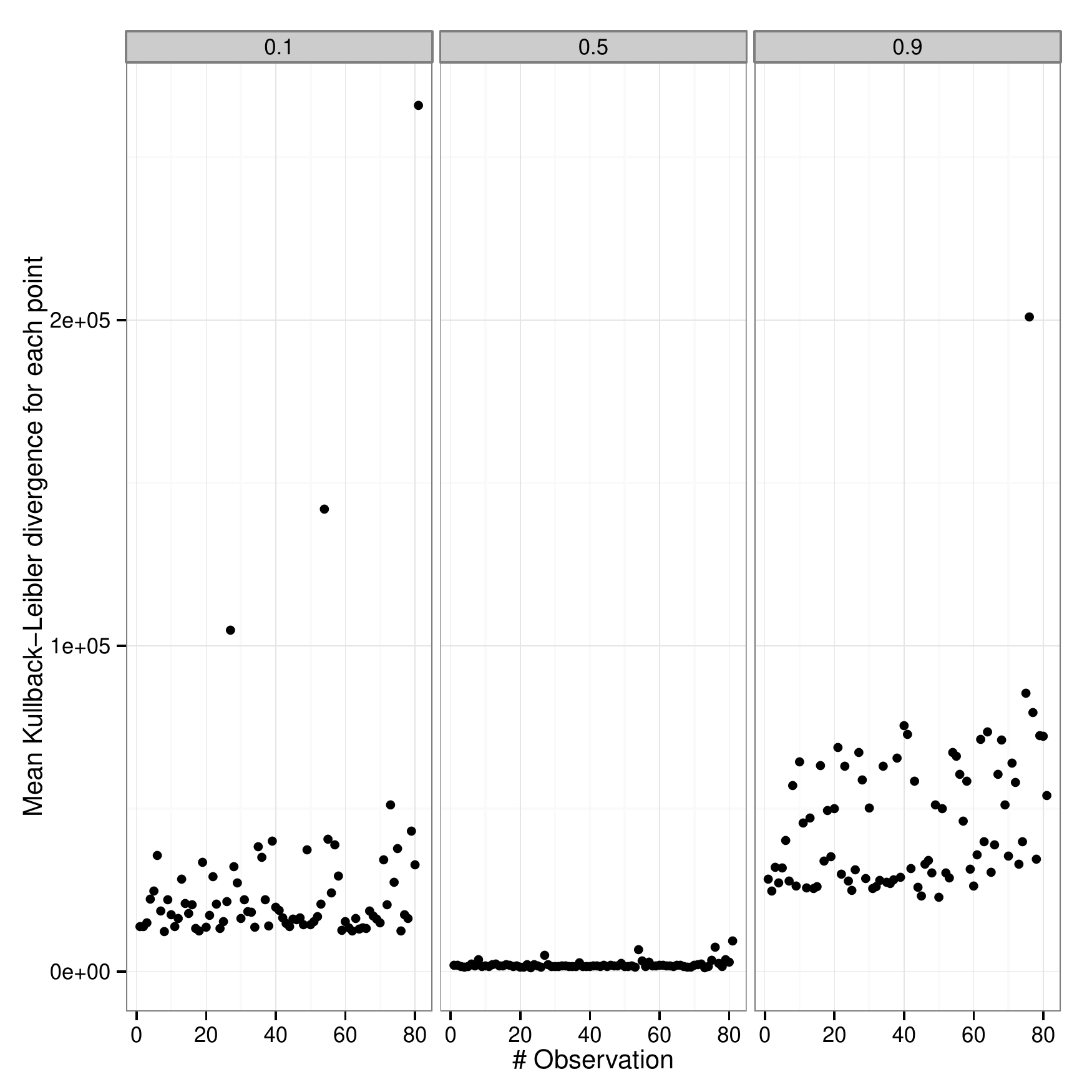}
\caption{$KL(f_i)$ for $\tau = \{0.1, 0.5, 0.9\}$, 
considering the model in \eqref{modelApplication}}
\label{figureKL}
\end{center}
\end{figure}

If we calculate the probability proposed in Section~\ref{outSection} for all 
observations we get Figure~\ref{figureOutliers}. And the Kullback-Leibler 
divergences are presented in Figure~\ref{figureKL}. Here we focus the attention 
on three quantiles, even though we analyzed the others quantiles, as only 
in these quantiles there were observations which are separated from 
the others in these plots. In the 0.1th quantile, these observations are 
\#27, \#54 and \#81, which are the three observations from Federal District,
in the three years that the data was collected. For quantile 0.9, the 
observation \#76 is the one most distant from the others, and it is 
about the state of Santa Catarina in the year 2010. Comparing figures 
\ref{figureOutliers} and \ref{figureKL}, we have the same pattern of 
observations which are detached from the others.

In the first case, those three observations from Federal District have high 
values of income per capita, with R\$917, R\$1,204, and R\$1,717 in the 
years 1991, 2000 and 2010, respectively. Ordering this variable in the 
sample we find that these values are number 8, 2, 1 in this list, respectively.
Also for years of education, these points present high values considering 
the data. Meanwhile, the effect of income per capita in the lower quantiles 
is estimated to be negative as shown in Figure~\ref{figureEstModel}, as it 
is the effect of years of education. On the other hand, their Gini indexes 
are among the highest in the dataset. Therefore, it is suiting that these 
observations are marked as outliers in the lower quantiles of the conditional 
distribution, given these unexpected results, as for all three it was likely
that they present small values for the Gini index.

Moreover, observation \#76 from the state of Santa Catarina, measured 
in 2010, has the lowest Gini in the sample, of 0.49. It is 
important to note that this observation presented a greater probability of being 
an outlier just in the higher quantiles. It can be argued that this observation 
should be considered an outlier since it presented the lower value of Gini in the 
sample despite happening in the year 2010, while the estimated coefficients for 
this dummy variable are positive for all quantiles, even though not significant 
for some quantiles. Besides that, this observation presents a big difference to next 
state in the sample, as the second lowest value is 0.53. Such a difference between 
two points is not seen in the entire sample, making it even more fitting this 
observation as an outlier.

These observations from two different states could be considered outliers in 
different parts of the conditional distribution of the Gini index, and this 
was only possible examining their latent variables in each quantile of interest,
as we propose here in this work.

\section{Final discussion}

Quantile regression models have become a great tool in the regression analysis 
framework given its flexibility in studying the conditional quantiles of the 
response variable. The Bayesian version of this model, taking into account the 
misspecified model assumption, is well established now with the 
asymmetric Laplace distribution and its mixture representation, which readily provides
a setup to identify possible outlying observations in the regression analysis, while
also controlling for the variance in the data with the $\sigma$ parameter. We 
showed how the posterior inference for $\sigma$ varies with $\tau$, and how it
could be vague when its value is fixed from the beginning. We also showed how 
the posterior distribution for each latent variable $v_i$ provides evidence
regarding observations that are too far apart from the others, which could be 
seen as outliers. We demonstrated these results with simulated examples to 
illustrate how this approach works, showing how when there are more than one 
outlier, they can affect the estimates differently for distinct quantiles.

In a real dataset, about Gini indexes in Brazilian states, we were able to find 
extreme observations from two different states that affected the quantile regression 
fits in different parts of the conditional distribution, one being in the lower 
quantiles and the other in the greater quantiles. This was only possible using 
our approach that gives attention to each quantile separately. It is important 
to note that in our method we are not checking whether this observation influences 
the regression models or not, as some diagnostics measures are concerned with, 
but we are more interested in identifying these most distant observations from 
the others, based on the posterior posterior distribution from their latent 
variable $v_i$, even though we did observe in the simulation studies that the 
outliers increased the bias in the quantile regression estimates. As a future 
study, case-deletion diagnostics for this type of model could be proposed, 
in addition to our approach.

\section*{Acknowledgements}

This research was supported by the Funda\c{c}\~ao de Amparo \`a Pesquisa do Estado 
de S\~ao Paulo (FAPESP) under Grants 2012/20267-9 and 2013/04419-6.

\bibliographystyle{imsart-nameyear}
\bibliography{Draft}

\end{document}